\documentclass[preprint2]{aastex62}%Apj 2columns
\pdfoutput=1 %for arXiv submissionok
\usepackage{amsmath,amstext}
\usepackage[T1]{fontenc}
\usepackage{apjfonts} 
\usepackage{comment}
\usepackage[figure,figure*]{hypcap}
\usepackage{multirow}
\usepackage{amsmath}
\usepackage{nccmath}
\usepackage{psfrag,graphicx,color}
\usepackage{soul}
\usepackage{lineno}
\usepackage{cancel}
\setstcolor{red}

\newcommand{\rahms}[4]{$#1^{\rm h}#2^{\rm m}#3\mbox{$^{\rm s}\mskip-7.6mu.\,$}#4$} %% \dechms{04}{14}{12}{9198} = RA en formato 04^h 14^m 12^s .9198
\newcommand{\decdms}[4]{$#1^{\circ}#2'#3\mbox{$''\mskip-7.6mu.\,$}#4$} %% \decdms{28}{12}{12}{199} = Dec en formato 28^o 12' 12'' .199

\shorttitle{VLA proper motions of PSR J1813--1749}
\shortauthors{Dzib \& Rodr\'{\i}guez}
\DeclareUnicodeCharacter{2212}{-}

\begin{document}

\title{Radio Proper Motions of the Energetic Pulsar PSR~J1813$-$1749}

\correspondingauthor{Sergio A. Dzib}
\email{sdzib@mpifr-bonn.mpg.de}

\author[0000-0001-6010-6200]{Sergio A. Dzib}
\affiliation{Max-Planck-Institut f\"ur Radioastronomie, Auf dem H\"ugel 69,
 D-53121 Bonn, Germany}
 \author[0000-0003-2737-5681]{Luis F. Rodr\'{\i}guez}
\affiliation{Instituto de Radioastronom\'{\i}a y Astrof\'{\i}sica, Universidad Nacional Aut\'onoma de M\'exico, Morelia, Michoac\'an 58089, Mexico}
\affiliation{Mesoamerican Centre for Theoretical Physics, Universidad Aut\'onoma de Chiapas, Tuxtla Gutiérrez, Chiapas 29050, Mexico}

\begin{abstract}
PSR J1813--1749 has peculiarities that make it a very interesting 
object of study. It is one of the most energetic and the most scattered 
pulsar known. It is associated with 
HESS J1813--178, one of the brightest and most compact TeV sources in the sky.
Recently, Ho et al. used archival X-ray Chandra observations separated by more 
than 10 years and determined
that the total proper motion of PSR J1813--1749 is $\sim66$~mas~yr$^{-1}$, 
corresponding to a velocity of $\sim$1900 km s$^{-1}$ for a distance of 6.2 kpc.
These results would imply that this pulsar is the fastest neutron star known 
in the Galaxy and, by estimating
the angular separation with respect to the center of the associated supernova remnant,
has an age of only $\sim300$ years, making it one of the youngest pulsars known.
Using archival high-angular-resolution VLA observations taken over 12 years we have estimated 
the radio proper motions of PSR~J1813--1748 to be much smaller: 
($\mu_{\alpha}\cdot\cos(\delta),\,\mu_\delta$)=($-5.0\pm3.7,\,-13.2\pm6.7$)~mas~yr$^{-1}$, or a total proper motion of $14.8\pm5.9$~mas~yr$^{-1}$.
The positions referenced against quasars make our results reliable. We conclude that PSR J1813--1749 is not a very fast moving source.
Its kinematic age using the new {total proper motion} is $\sim1350$~years.
This age is consistent within a factor of a few with the characteristic age of the pulsar and with the age
estimated from the broadband spectral energy distribution of HESS J1813--178, {as well as the age of the associated SNR}. 

\end{abstract}


\nopagebreak
\section{Introduction}

\begin{figure*}[!ht]
   \centering
\includegraphics[height=0.5\linewidth, trim=0 0 0 0, clip]{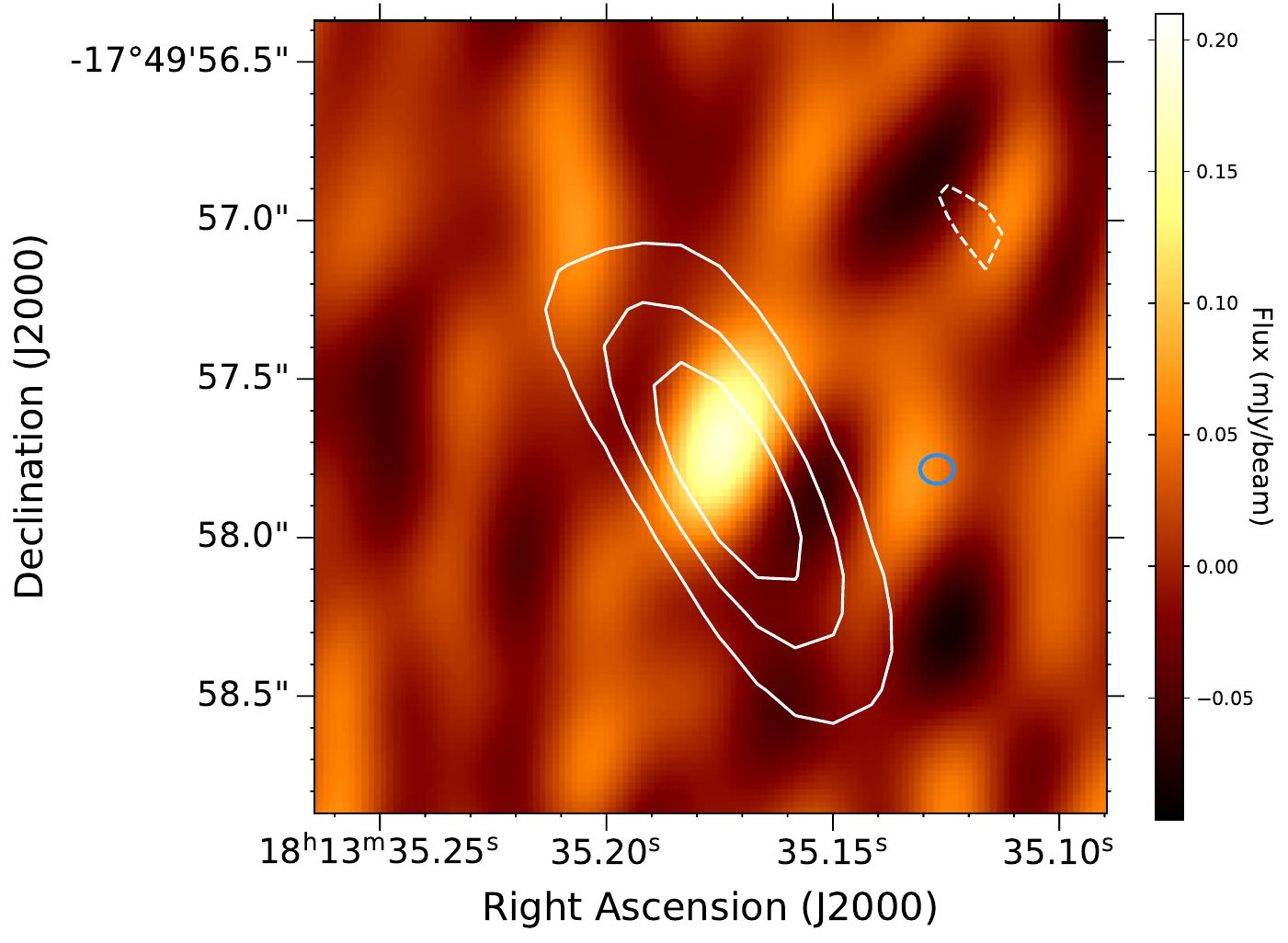}

   \caption{{\it Background:} VLA image of PSR J1813--1749 as observed in February 2006. {\it Contours:} PSR J1813--1749 as observed in December 2017 at X-band. Contour levels are -3, 3, 6, and 9 times 7\,$\mu$Jy~beam$^{-1}$, the noise level on this epoch. {The blue} ellipse indicates the expected position of the radio source in December 2017, following the proper motion measured by \citet{ho2020}, the ellipse semi-major axis sizes consider the propagated errors.}
   \label{fig:SNR}
\end{figure*}

PSR J1813--1749 (=CXOU J181335.1$-$174957) is the most 
scattered and the second most energetic pulsar in the 
Milky Way \citep{halpern2012,camilo2021}.
First discovered and characterized using Chandra X-ray 
observations \citep{GH2009, halpern2012}, it has a spin-down 
rate of $\dot{P}=1.265\times10^{-13}$, corresponding 
to a spin-down luminosity of 
$\dot{E}= 5.6\times10^{37}$~erg~s$^{-1}$, values only below 
those measured for the Crab pulsar \citep[e.g.,][]{halpern2012}.

First attempts to detect the radio pulsed emission 
at low frequencies (1--2~GHz) from this pulsar failed 
\citep{helfand2007,halpern2012,dzib2018}. Recently, \citet{camilo2021}
finally detected the pulsed radio emission at higher frequencies 
(4--10~GHz) and showed that the pulses are highly scattered.
The fact that the scattering is more severe at
lower frequencies probably explains 
the early failed attempts to detect the radio pulsed emission.
\citet{camilo2021} show that the pulsed emission is consistent 
with the radio continuum source detected by \citet{dzib2010}
and \citet{dzib2018} with the Karl G. Jansky Very Large Array (VLA) 
at similar frequencies. Based on the high column density
at X-rays and the large dispersion measure, \citet{camilo2021} 
also place a lower limit to its distance of 6.2 kpc, that could 
be as large as 12~kpc.

The young and relatively compact ($\sim2'$ diameter) shell-type radio 
supernova remnant (SNR) G12.82-0.02 \citep{brogan2005} and pulsar 
wind nebula (PWN) observed at X-rays 
\citep[][]{helfand2007,funk2007,GH2009} have been associated to 
PSR J1813--1749. SNR G12.82--0.02 and the PWN are associated with 
one of the brightest and most compact objects discovered by the HESS 
Galactic Plane Survey \citep{aharonian2005}, the TeV source 
HESS J1813--178. This HESS source has been associated with
continuum high-energy emission from X-rays to gamma rays 
\citep{reimer2008,ubertini2005,abdo2009,albert2006}.

Recently, using archival X-ray observations, \citet{ho2020} 
determined large proper motions for PSR J1813--1749 of 
($\mu_{\alpha}\cdot\cos(\delta),\,\mu_{\delta}$)= 
(${-64\pm9},\,-14\pm7$)~mas~yr$^{-1}$. As the pulsar is at an angular
distance of $\sim20''$ from the center of the SNR G12.82--0.02 the large proper motions  
would indicate a young age of around  300~years, making it one of the youngest 
pulsars known. This age, { while consistent at the lower end of the age range of
285 to 2500 years for SNR G12.82--0.02 \citep{brogan2005}, is however}  in  conflict  with the age
estimated for HESS~J1813--178 of 2500~years \citep{zhu2018} and the 
characteristic age of the pulsar of 5600 years \citep{halpern2012}.
As discussed by
\citet{camilo2021}, { the total proper motion of $\sim66$~mas yr$^{-1}$} would imply a tangential
velocity of the order of 2,000~km~s$^{-1}$ at 6.2~kpc, the
lower limit of the distance. This velocity is larger than that of any well-measured velocity for a neutron star (cf., Deller et al. 2019).
The {total} proper motion of PSR~J1814-1749  is an interesting subject to study, 
and in this paper we present the proper motions measured with archival
high angular resolution observations taken with the VLA by us and by other groups.

\section{VLA Observations}

For our astrometric study we looked for VLA observations with high 
angular resolution. The VLA provides the finest angular resolution
in its most extended configurations A and B. We also restricted our 
search to the C-band (4--8 GHz) and X-band (8--12\,GHz) which provide 
the best sensitivity, and where PSR J1813--1749 has been previously 
detected. For the best astrometry it is also recommended  
that the observations are phase referenced to the same quasar (gain calibrator),
as this provides nearly absolute astrometry. 

We found two observational campaigns with all the above criteria and 
where the target source has been detected. These observations have 
been previously reported by \citet{dzib2010} and \citet{dzib2018}.
The first is one observation done in 2006 with the historical VLA 
at 4.86~GHz, using the A-configuration, under project AL673. The second, includes a 
series of 12 observations, 5 centered at a main frequency of 6.0\,GHz 
and 7 at 10.0\,GHz {that were made as part of project 17B-028}. The observations were done in the B-configuration
covering the period from September 2017 to February 2018. We also found 
a third observational campaign done in October 2012 {under project 12B-278}, at the mean frequency
of 9.0 GHz, using the VLA in its A configuration. However, this third 
campaign used a different gain calibrator. In section 2.1 we discuss how we 
corrected for this limitation. All observations were calibrated
and imaged using the CASA software. Positions were determined from the image
using the CASA task {\it imfit}. Fluxes and other emission properties were
already given and discussed by \citet{dzib2010} and \citet{dzib2018}, and 
in this work we focus on the astrometry. Basic properties of the image and 
position of the target source over time are listed in Table~\ref{tab:vlapsr}

\begin{table*}
\small
\centering
\renewcommand{\arraystretch}{0.85}
\caption{Image results, and positions of PSR~J1813$-$1749.}
\begin{tabular}{lcccccccc}\hline\hline
Epoch       & $\nu$ & VLA&{Gain} & Beam  size                    & Noise               & R.A. (J2000)    &Dec. (J2000)      \\
(yyyy.mm.dd)&(GHz)&Conf.&Calibrator&[$''\times'';\,\, ^{\circ}$]&($\mu$Jy bm$^{-1}$)  & (18$^{h}13^{m}$)&(--17$^\circ49'$) \\
(1)&(2)&(3)&(4)&(5)&(6)&(7)&(8)\\
\hline
2006 Feb. 25 & 4.8 &  A & J1811-2055&$0.77\times0.41;$ --9 & 19&  35\rlap{.}$^{\rm s}$177 $\pm0\rlap{.}^{\rm s}$003  &  57\rlap{.}$ ''62\pm0\rlap{.}''$08   \\
2012 Oct. 05$^{a}$ & 9.0 &  A & J1733-1304&$0.45\times0.22;$  19  &  11&  35\rlap{.}$^{\rm s}$174 $\pm0\rlap{.}^{\rm s}$009  &  57\rlap{.}$ ''56\pm0\rlap{.}''$13\\ 
%2012 Oct. 06 & 5.5 &  A & J1733-1304&$0.52\times0.26;$ 25  & 13& \\ 
2017 Sept. 18& 6.0 &  B & J1811-2055&$1.51\times0.74;$ 15  & 8 &  35\rlap{.}$^{\rm s}$172 $\pm0\rlap{.}^{\rm s}$003  &  57\rlap{.}$ ''85\pm0\rlap{.}''$08\\ 
2017 Sept. 18& 10.0&  B & J1811-2055&$0.95\times0.46;$ 18  & 6 &  35\rlap{.}$^{\rm s}$176 $\pm0\rlap{.}^{\rm s}$002  &  57\rlap{.}$ ''72\pm0\rlap{.}''$06\\ 
2017 Dec. 11 & 6.0 &  B & J1811-2055&$1.74\times0.77;$ 33  & 9 & 35\rlap{.}$^{\rm s}$173 $\pm0\rlap{.}^{\rm s}$003   &  57\rlap{.}$ ''74\pm0\rlap{.}''$05\\ 
2017 Dec. 11 & 10.0&  B & J1811-2055&$1.18\times0.49;$ 35  & 7 &  35\rlap{.}$^{\rm s}$175 $\pm0\rlap{.}^{\rm s}$002  &  57\rlap{.}$ ''77\pm0\rlap{.}''$05 \\ 
2018 Jan. 8  & 6.0 &  B & J1811-2055&$1.49\times0.77;$ 23  & 7 & 35\rlap{.}$^{\rm s}$172 $\pm0\rlap{.}^{\rm s}$002   &  57\rlap{.}$ ''78\pm0\rlap{.}''$05\\ 
2018 Jan. 8  & 10.0&  B & J1811-2055&$0.98\times0.47;$ 26  & 7 & 35\rlap{.}$^{\rm s}$178 $\pm0\rlap{.}^{\rm s}$002   &  57\rlap{.}$ ''73\pm0\rlap{.}''$05\\ 
2018 Jan. 13 & 6.0 &  B & J1811-2055&$1.76\times0.76;$ 32  & 8 & 35\rlap{.}$^{\rm s}$174 $\pm0\rlap{.}^{\rm s}$003   &  57\rlap{.}$ ''75\pm0\rlap{.}''$06\\ 
2018 Jan. 13 & 10.0&  B & J1811-2055&$1.19\times0.47;$ 35  & 8 & 35\rlap{.}$^{\rm s}$175 $\pm0\rlap{.}^{\rm s}$004   &  57\rlap{.}$ ''64\pm0\rlap{.}''$09\\ 
2018 Jan. 21 & 10.0&  B & J1811-2055&$1.40\times0.51;$ 31  & 7 & 35\rlap{.}$^{\rm s}$176 $\pm0\rlap{.}^{\rm s}$004   &  57\rlap{.}$ ''88\pm0\rlap{.}''$09\\ 
2018 Jan. 28 & 6.0 &  B & J1811-2055&$1.44\times0.77;$ 16  & 7 & 35\rlap{.}$^{\rm s}$171 $\pm0\rlap{.}^{\rm s}$001   &  57\rlap{.}$ ''80\pm0\rlap{.}''$04\\ 
2018 Jan. 28 & 10.0&  B & J1811-2055&$0.93\times0.48;$ 21  & 6 & 35\rlap{.}$^{\rm s}$174 $\pm0\rlap{.}^{\rm s}$002   &  57\rlap{.}$ ''72\pm0\rlap{.}''$04\\ 
2018 Feb. 4  & 10.0&  BnA& J1811-2055&$1.16\times0.25;$ -57& 8 & 35\rlap{.}$^{\rm s}$171 $\pm0\rlap{.}^{\rm s}$006   &  57\rlap{.}$ ''70\pm0\rlap{.}''$05\\ 
\hline\hline
\label{tab:vlapsr}
\end{tabular}

\noindent {\footnotesize $^{a}$ Data averaged from the two observed dates: Oct. 4 and Oct. 6 2012. The given position for this epoch
 is already corrected for the systematic offset discussed in sub-section 2.1.}
\end{table*}

\subsection{Astrometric correction for the 2012 observations}
The observations of 2006 and 2017-18 were made with the same gain calibrator  (J1811$-$2055). This typically assures an astrometric precision of order $\simeq 0\rlap.{''}01$ { for the case of observations made at centimeter wavelengths in the A-configuration \citep{boboltz2007,perreault2019}. We can estimate from our data the expected precision as follows. We will use the ten determinations of position given in Table 1 for the epochs between 2017 December 11 and 2018 February 4. Over this brief period of time we do not expect significant proper motions. The positional error scales linearly with the angular size of the synthesized beam. Since all these observations were made in the B configuration we expect positional errors about three times larger than in A configuration, that is, about 30 mas. Furthermore, since this is a southern source, we expect the beam size to be about twice bigger in declination that in right ascension (see values in Table~\ref{tab:vlapsr}). We indeed find that the root-mean-square errors in right ascension and declination for these ten observations are 31.8 and 63.8 mas, respectively, 
approximately as expected. We conclude that the positional error is given approximately by the angular size of the 
synthesized beam over 20, that is $\theta$/20.} 

We were very interested in using the 2012 observations of project 12B$-$278 to  derive a measurement intermediate in time. However, these observations used a different gain calibrator (J1733$-$1304) and this can introduce systematic position errors of order $\simeq 0\rlap.{''}1$ or more unless a correction is applied to the positions. The 12B$-$278 observations were made during 2012 October 04 and 06 and we concatenated the data for an average epoch of 2012.762. The observations were made in the A configuration in band X (8.0-10.0 GHz), with 16 spectral windows of 128 MHz width each.

\begin{table*}
\small
\centering
\renewcommand{\arraystretch}{0.85}
\caption{Gaia sources used to correct the astrometry of the 12B-078 positions$^a$. }
\begin{tabular}{lcccccccc}\hline\hline
Source       & Type & R.A. (J2000)    &Dec. (J2000)& $\mu_{\alpha}\cdot\cos{(\delta)}$ & $\mu_{\delta}$      \\
(1)&(2)&(3)&(4)&(5)&(6)\\
\hline
CXOUJ181314.2-175343 &   Wolf-Rayet & \rahms{18}{13}{14}{200} &  
\decdms{-17}{53}{43}{46} &     $-0.91\pm0.07$ & $-1.99\pm0.05$\\
2MASSJ18131908-1752585  &  O8-O9If&   \rahms{18}{13}{19}{079}&   
\decdms{-17}{52}{58}{49}&     $-1.11\pm0.13$ & $-1.81\pm0.10$\\
$[$MFD2008$]$ 15            &   LBV &     \rahms{18}{13}{20}{984} &  
\decdms{-17}{49}{47}{02} &     $-0.57\pm0.07$ & $-1.98\pm0.06$\\
CXOUJ181322.4-175350    &Wolf-Rayet & \rahms{18}{13}{22}{492} & 
\decdms{-17}{53}{50}{32}  &    $-0.78\pm0.07$ & $-2.08\pm0.05$\\
UCAC2 25154559          &B0-B3 star&  \rahms{18}{13}{24}{432} & 
\decdms{-17}{52}{56}{78} &     $-0.85\pm0.06$ & $-2.16\pm0.05$\\ 
\hline\hline
\label{tab:gaia}
\end{tabular}

\noindent {\footnotesize $^a$ Positions and proper motions are from the Gaia eDR3 data release \citep{gaia2021}. The positions of Gaia are accurate to 1 mas or better. The proper motions
are given in mas yr$^{-1}$. }
\end{table*}

The final images of the two epochs of project 12B-278 and the 5 C-band epochs
of project 17B-028
were compared. In addition to the source associated with PSR J1813$-$1749
we found five compact sources in common { that are also detected in the Gaia survey \citep{gaia2016,gaia2021}}. The highly accurate Gaia positions and proper motions of these
five sources are given in Table~\ref{tab:gaia}. We corrected the radio positions of these sources with the Gaia proper motions and used them to determine a systematic offset between the positions obtained in the 17B-028 and 12B-278 projects. This offset (17B-028 -- 12B-78) is 
${\Delta RA = 0\rlap.^s0110 \pm  0\rlap.^s0091}$; ${\Delta 
DEC = -0\rlap.^{''}035 \pm 0\rlap.^{''}125}$. After adding this offset 
to the positions of the 12B$-$278 project, we obtain a final position 
for PSR J1813$-$1749 at this epoch:

\begin{align*}
RA(J2000) = 18^h~13^m~35\rlap.^s180 \pm 0\rlap.^s009;\\
DEC(J2000) = -17^\circ~49' 57\rlap.^{''}52 \pm 0\rlap.^{''}13.
\end{align*}
We note that the error in the final position is dominated by the offset correction applied. %The positions of PSR J1813--1749 as a function of time are shown in Figure~\ref{fig:FGaC}.

\section{Results}

\begin{figure}
    \centering
    \includegraphics[width=0.47\textwidth, trim= 0 0 0 0,clip, angle=0] {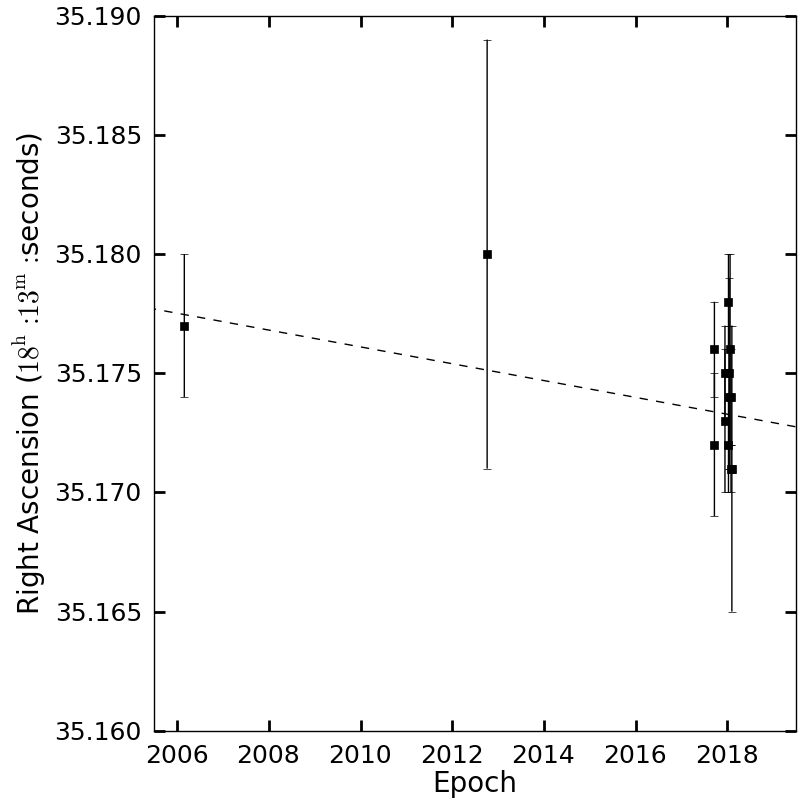}\\ 
    \includegraphics[width=0.47\textwidth, trim= 0 0 0 0,clip, angle=0] {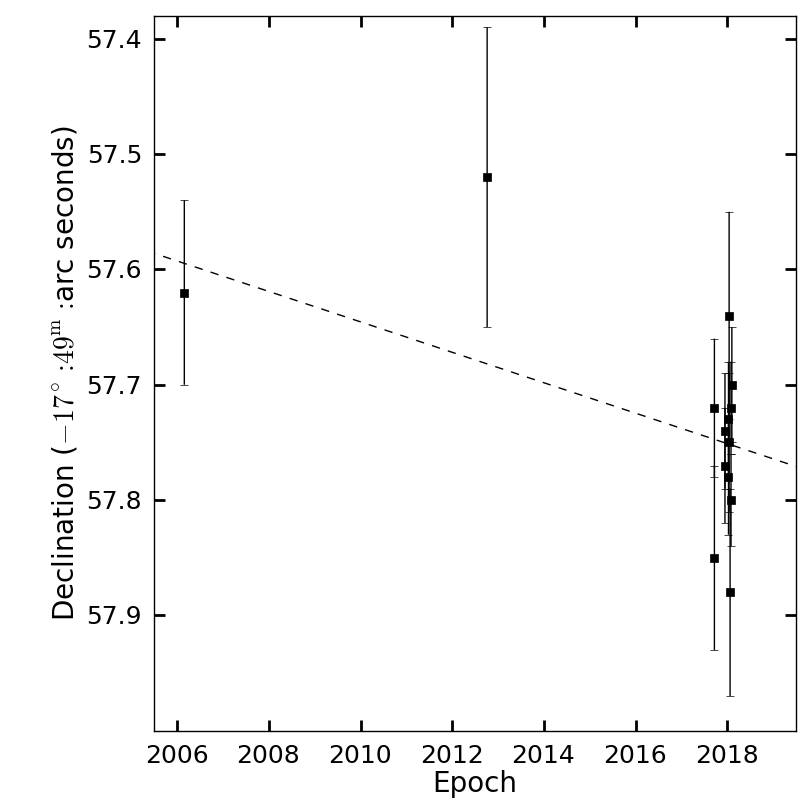}
    \caption{Right ascension (top) and declination (bottom) of the PSR J1813--1749 radio emission
    as a function of time.  The dashed lines are least-squares linear fits to the data. The parameters of the fits are given in the text.}
    \label{fig:PM}
\end{figure}

We have performed linear least square fits to the positions of PSR J1813--1749, listed 
in Table~\ref{tab:vlapsr}, to determine its proper motions. 
The {values obtained} are ($\mu_{\alpha}\cdot\cos{\delta},\,\mu_\delta$)=(${-5.0\pm3.7,\,-13.2\pm6.7}$) mas~yr$^{-1}$.
The positions of PSR~1813--1749 as a function of time, and the best fit
to its motion are shown in Figure~\ref{fig:PM}.

\section{Discussion and Conclusions}
 
{The proper motions of PSR J1813--1749 determined from the X-ray observations
are ($\mu_{\alpha, X-rays}\cdot\cos{\delta},\,\mu_{\delta, X-rays}$)= 
($-64\pm9,\,-14\pm7$) mas~yr$^{-1}$. By comparing them with our results,
we clearly notice a difference in the R.A. being at radio significantly smaller.}
Proper motions measured at radio have as major advantage over 
the X-ray measurements that the positions are registered against 
highly accurate positions measured for quasars. The observations 
presented in this work also have a somewhat larger 
time baseline, 12 years, than that of the X-ray observations presented by 
\citet{ho2020} of 10 years.  This difference does not seem too significant, but it should be
emphasized that the accuracy of proper motion determinations improves as the time
interval to the 3/2 power \citep{dzib2017}. The motions reported at X-rays would be 
evident, between the first and last radio observations, in positions 
offsets of --0\rlap{.}$''$80=0\rlap{.}$^{\rm s}$05 and --0\rlap{.}$''$17 
in R.A. and Dec., respectively. {The offset in
R.A. is clearly not present, see also Fig.\ref{fig:SNR}.}

The large proper motion has also been questioned given 
the strong implications for the nature of the pulsar since it 
 implies a young kinematic age and a fast tangential velocity larger
than any other known pulsar \citep[see also the discussion by][]{camilo2021}.

Both radio and X-ray emission are tracing the pulsar itself or material very close to it. However,
given the discussion above, the measured motions at radio frequencies 
appear to be more reliable. We believe that most of the position shift in the X-ray image
could be due to a change in the brightness structure of the PWN very near the pulsar,
a possibility mentioned by \citet{ho2020}. {Such structure changes have been observed in the Crab pulsar Nebula \citep{weisskopf2011}.}

%\subsection{Implication of the motions}

The total proper motion of PSR J1813--1749 from the radio is ${14.8\pm5.9}$\,mas\,year$^{-1}$. 
The lower limit of the distance to PSR J1813--1749 is 6.2\,kpc and can be 
as large as 12.0\,kpc. Then, the tangential velocity ranges from 
${435\pm174}$\,km\,s$^{-1}$ to  ${842\pm336}$\,km\,s$^{-1}$. These velocities are 
in the range of velocities estimated for other pulsars \citep[i.e.,][]{deller2019}. It should be noted that
the error in the proper motion is large enough to accept a stationary pulsar as a possible solution.

As noted by \citet{ho2020}, PSR J1813--1749 is offset about $20''$ from the
center of the SNR G12.82--0.02 \citep[see also Figure 1 in ][]{dzib2018}. 
To reach this shift the kinematic age of the pulsar is 
{1351}${^{+896}_{-385}}$ years. This age discards that this is a very young pulsar 
and it is in better agreement with the ages estimated for the pulsar
of 5600 years \citep{halpern2012} and for HESS J1813--178 of 2500 years 
\citep{zhu2018} { and for SNR G12.82--0.02 \citep{brogan2005}}. 

To calculate the kinematic age we have assumed that the original position of the exploding star
was at the geometric center of the supernova remnant. However, some SNRs have shown nonuniform 
expansion \citep[e.g.,][]{borkowski2014} and the geometric center of the present-day 
structure does not necessarily coincide with the center of the explosion.

VLA observations have proven to be an excellent tool to determine proper motions,
and in the case of PSR J1813--1749, are at the moment the best option.
Even VLBI observations will have difficulties  measuring a value for this source. 
The angular size of the radio source is estimated to be 0\rlap{.}$''$034
\citep{camilo2021} due to broadening from {plasma} scattering. The angular resolution
of VLBA observations at 5 GHz is $\sim0\rlap{.}''004$, {so} the source will
be resolved. 
The total flux density of the source is $\sim100\,\mu$Jy \citep{dzib2018}; 
if resolved very little flux density will fall in a synthesized beam and the emission will be 
hard to detect
with standard VLBA observations. Furthermore, astrometry of resolved sources is
problematic. To better constrain the proper motion of 
PSR J1813--1749, future VLA observations, as those presented here, will be 
required.

%\section{Conclusions}

\acknowledgements 
%{\small
We thank an anonymous referee for valuable comments. L.F.R.  acknowledges the financial support of DGAPA, UNAM (project IN108920), and CONACyT, M\'exico. 
The National Radio Astronomy Observatory is a facility of the National Science Foundation operated under 
cooperative agreement by Associated Universities, Inc. 

\facilities{VLA} 
\software{ CASA \citep{mcmullin2007}.}

\bibliographystyle{aa}

\end{document}